\documentclass[preprint,aps,nofootinbib,11pt]{revtex4-1}
\usepackage{geometry}                
\usepackage{amsmath,amssymb,amsfonts,dcolumn,color,graphicx,graphics,latexsym,placeins,epsfig,tikz}
\usetikzlibrary{arrows,shapes}
\usepackage{subfigure,rotating,bm,mathrsfs}

\newcommand{\be}{\begin{equation}}
\newcommand{\ee}{\end{equation}}
\newcommand{\ba}{\begin{eqnarray}}
\newcommand{\ea}{\end{eqnarray}}

\usepackage[pagebackref=false, colorlinks=true]{hyperref}
\definecolor{redish}{rgb}{0.7,0.2,0.0}  
\definecolor{bluish}{rgb}{0.2,0.5,0.8}

\hypersetup{linkcolor=redish,          
                  citecolor=blue,        
                  filecolor=magenta,      
                  urlcolor=blue}          

\begin{document}
\author{Rajibul Shaikh}
\email{rshaikh@iitk.ac.in}
\author{Pritam Banerjee}
\email{bpritam@iitk.ac.in}
\author{Suvankar Paul}
\email{svnkr@iitk.ac.in}
\author{Tapobrata Sarkar}
\email{tapo@iitk.ac.in}
\affiliation{Department of Physics, \\ Indian Institute of Technology, \\ Kanpur 208016, India}
\title{Analytical approach to strong gravitational lensing from ultracompact objects}

\begin{abstract}
Strong gravitational lensing from black holes results in the formation of relativistic images, in particular,
relativistic Einstein rings. For objects with event horizons, the radius of the unstable light ring (photon sphere) is the lowest radius at which
a relativistic image might be formed. For horizonless ultracompact objects, additional relativistic images and rings
can form inside this radius. In this paper, we provide an analytical approach to deal with strong gravitational 
lensing from such ultracompact objects, which is substantially different from the black hole cases, first reported
by Bozza. Here, our analysis indicates that the angular separations and magnifications of 
relativistic images inside the unstable light ring (photon sphere) might be several orders of magnitude higher compared to the 
ones outside it. This indicates fundamental differences in the nature of strong gravitational lensing from black holes
and ultracompact objects. 

\end{abstract}

\maketitle

\section{Introduction}

Bending of light in a gravitational field, known as gravitational lensing \cite{Weinberg}, continues to be an important 
focus of research, a century after it was first experimentally observed. In the context of astronomical observations, weak
gravitational lensing has played a significant role in our understanding of galactic constituents. However, 
in situations involving black holes or compact astrophysical objects, one naturally invokes bending of light due 
to strong gravity \cite{BozzaReview}. It is of fundamental importance to study these issues further, in the light of the
recent efforts to obtain black hole images by the Event Horizon Telescope \cite{EHT}. 

The photon sphere (or unstable light ring) is ubiquitous in this context \citep{BozzaReview,SL1,SL2,SL3}, and is proposed
as one of the main diagnostic tools for mapping the black hole event horizon. In natural units ($G = c = 1$),
the location of the photon sphere in Schwarzschild coordinates is at $r = 3M$ for the Schwarzschild black hole, where gravity becomes strong enough for
a photon to have an unstable circular orbit so that a small perturbation can cause the photon to be either absorbed by the 
black hole or sent off to a faraway observer. In the second case, when the observer, the source and the lens are in alignment, the photon sphere results in relativistic Einstein rings.

In recent years, horizonless objects have attracted much attention for several reasons (see \cite{UCO1,UCO2,UCO3,UCO4,UCO5} and references therein). There has also been a lot of effort on whether or to what extent one can distinguish such horizonless compact objects from black holes. In light of this, gravitational lensing and its various aspects by different horizonless objects such as wormholes \cite{WL1,WL2,WL3,WL4,WL5,WL6,WL7,WL8,WL9,WL10,WL11,WL12,WL13,WL14,WL15,WL16, WL17,WL18,WL19,WL20,WL21,WL22}, naked singularities \cite{NL1,NL2,NL3,NL4,NL5,NL6,NL7,NL8}, Bosonic stars \cite{Cunha1}, compact object with arbitrary quadrupole moment \cite{CO}, gravastar \cite{gravastar} etc. have been analyzed. However, somewhat less studied in the lensing literature is the role of the antiphoton sphere (stable light ring), which invariably arises in the
study of ultra compact objects (UCOs), which have an unstable light ring but no event horizon. 
This is the radius at which the photons can travel in a stable circular orbit.\footnote{Throughout
this work, we study the motion of photons dictated purely by geometry. Interaction between light and matter in
the interior of compact objects is a much more subtle issue and is not considered here.}
The study of UCOs is fast gaining popularity as a possible laboratory for testing gravitational lensing in 
astrophysical scenarios. In \cite{Kunha} the authors showed that light rings in UCOs must appear in pairs (see \cite{Hod}
for a possible counterexample). It is known that lensing features from UCOs can be vastly different
compared to those from objects with horizons \cite{Cunha1},\cite{Cunha2}. 
Such features (if observed) can distinguish between UCOs and black holes. 

\begin{figure}[h]
\centering
\centerline{\includegraphics[scale=0.7]{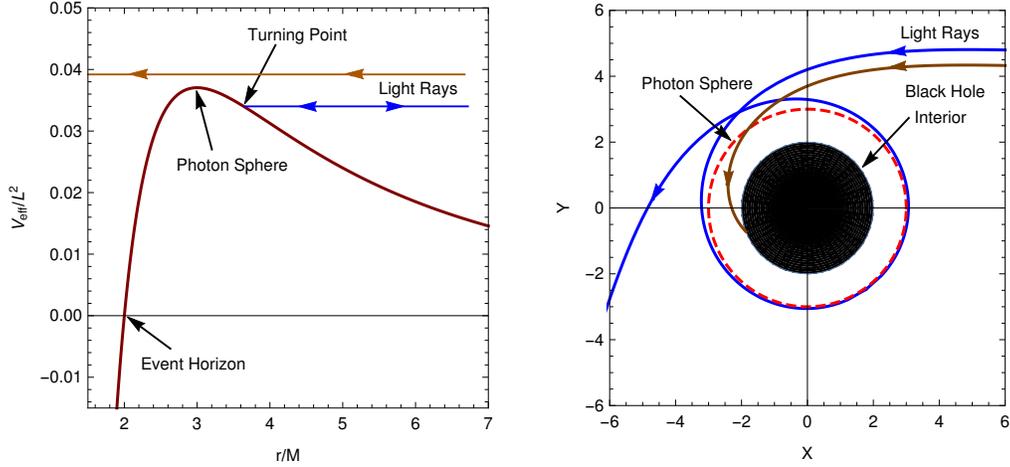}}
\caption{Strong lensing in a black hole space-time. We have used the Schwarzschild black hole for illustration.} 
\label{fig1}
\end{figure}

\begin{figure}
\centering
\centerline{\includegraphics[scale=0.7]{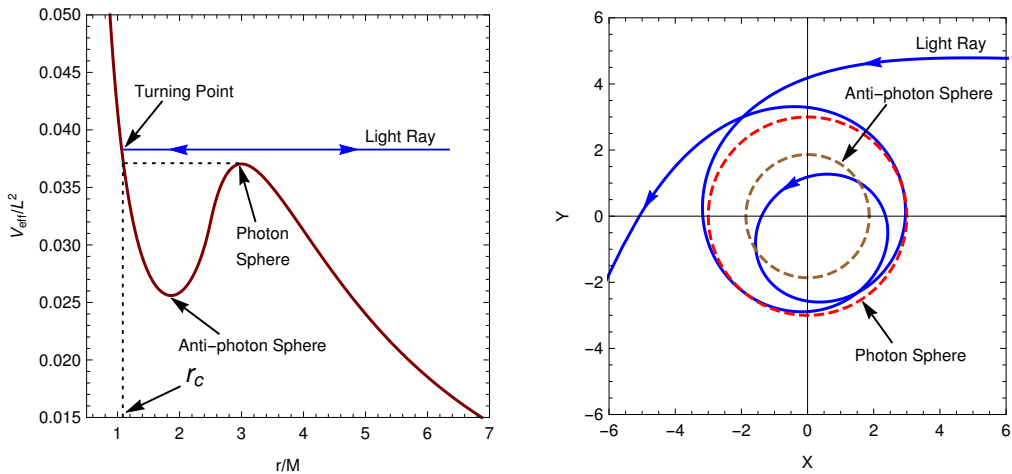}}
\caption{Additional images in strong lensing by an ultracompact object. For illustration, we have used Schwarzschild interior solution due to Synge with matching radius $R=2.5 M$, $M$ being the mass [see Eq. (\ref{eq:synge})]. Here, $r_c$ is the point where the effective potential (in units of angular momentum squared) has the same height as that at the photon sphere.} 
\label{fig2}
\end{figure}

Whereas previous studies on lensing from UCOs have been numerical \cite{Cunha1}, we perform
an analytic study here. 
The main idea that we develop in this paper is as follows. Figure \ref{fig1} qualitatively 
depicts the effective potential of geodesic motion for photons [in units of its angular momentum squared, see Eq. (\ref{Veffec})] 
in a static, spherically symmetric space-time corresponding to a black hole. For a certain impact parameter, 
photons that approach the black hole from a source at infinity will be trapped at the location of the photon sphere, where it will
undergo multiple rotations, until due to a small perturbation, it either escapes to infinity or is absorbed by the black hole. 
For UCOs, apart from the photons that escape
to infinity from the photon sphere, there is an extra set of images.
Namely, a photon that crosses the radius of the
photon sphere might be reflected at an internal point, whence it comes back to the photon sphere and can then
escape to an observer at infinity. This is depicted in Fig. \ref{fig2}. 
The two situations are fundamentally different. The first case  has been considered
in details by Bozza in \cite{SL3} and this analysis has recently been refined by Tsukamoto in \cite{Tsu1}. In this paper,
we focus on the second situation, which calls for a different analysis.


The work of Bozza and Tsukamoto assumes a turning point for a photon (coming from source) at a radial distance
greater than the photon sphere. The strong deflection limit is then obtained by taking the limit in which this turning
point tends to the photon sphere radius. Such a computation is clearly not suitable when one analyzes the 
photons that are reflected at a point inside the photon sphere. From Fig. \ref{fig2}, we see that in this case, to compute
the positions of relativistic images, one has to consider the turning point of a photon inside the antiphoton sphere at a point at which the
effective potential has nearly the same value as that at the photon sphere. As we show in sequel, this completely 
changes the analytical formulas presented in \cite{SL3}, \cite{Tsu1} and reveals important new features about gravitational 
lensing from UCOs. Here, we point out that a similar scenario has been considered in \cite{mandar} where the authors have obtained an analytic expression for the bending angle in the strong deflection limit in the presence of an antiphoton sphere. However, their result is obtained for a specific spacetime geometry, namely the Majumdar-Papapetrou dihole spacetime. Our focus here, however, is to obtain an analytic expression for the bending angle in the strong deflection limit as well as angular separations and magnifications of the relativistic images formed due to the presence of an antiphoton sphere of an arbitrary spacetime geometry representing an UCO.

It is known that for black holes, relativistic images are always formed at radii greater than that of the photon sphere, and that UCOs indicate a different result, namely that such images might be formed inside
the radius of the photon sphere. In fact, our results indicate that the images inside the photon sphere might in principle be easier to detect
than the ones outside it, as the angular separation and the magnification of the former can even be 2 
orders of magnitude greater than the latter. This is a novel feature of gravitational lensing from UCOs compared to
the black hole case. 

It should be pointed out that we are assuming here that a spherically symmetric and static black hole does not have an antiphoton sphere outside its event horizon. Indeed, if this was the case, then such a black hole might mimic the results presented here. Although we are not aware of a rigorous proof of this statement, our assumption is strongly motivated by the fact that, to the best of our knowledge, such a situation is not currently known either in the context of general relativistic black holes or those that appear in modified gravity. If the exterior geometry of a black hole possesses an antiphoton sphere, then, in addition to this, there must exist in this geometry two more photon spheres, since the photon effective potential vanishes both at the event horizon and at spatial infinity, and is positive. As of now, we are not aware of such a black hole solution that will have three such surfaces where photons can have circular (stable or unstable) orbits. Our focus in this work is on UCOs, which possess both photon and antiphoton spheres.

This paper is organized as follows. In the next section, we briefly summarize known results on generic deflection angles
of photons due to lensing by a gravitating object. In Sec. \ref{sec:3}, we study such lensing behavior in the strong deflection
limit. Here, we first recall known results due to lensing by a black hole in Sec. \ref{subsec:3a}. In Sec. \ref{subsec:3b}, the effect of 
an antiphoton sphere (stable light ring) in the gravitational lensing of photons is established.  
In Sec. \ref{sec:4}, we first recall the definitions of observables in gravitational lensing and obtain their analytic expressions for the relativistic images formed inside the photon sphere and tabulate
our results for the different geometries we consider. Finally, Sec. \ref{sec:5} ends with discussions on our
results and some broad conclusions. 

\section{Deflection angle for static, spherically symmetric space-times}
\label{sec:2}

In this section, we briefly recapitulate the necessary details about the deflection angle of light in an arbitrary
static, spherically symmetric space-time, with the line element given by 
\begin{equation}
ds^{2}=-A(r)dt^{2}+B(r)dr^{2}+C(r)(d\theta^{2}+\sin^{2}\theta d\phi^{2}),
\label{genmetric}
\end{equation}
where $A(r)$, $B(r)$, and $C(r)$ satisfy the asymptotically flat conditions
\begin{eqnarray}
\lim_{r\rightarrow \infty} A(r) = 1~,~\lim_{r\rightarrow \infty} B(r) = 1~,~\lim_{r\rightarrow \infty} C(r) = r^{2}.
\end{eqnarray}
For simplicity, we restrict ourselves to $\theta=\pi/2$. Because of the spherical 
symmetry, the same results can be applied to all $\theta$. Therefore, the 
Lagrangian describing the motion of a photon in the $\theta=\pi/2$ plane of 
the space-time geometry of Eq. (\ref{genmetric}) is given by
\begin{equation}
2\mathcal{L}=-A(r)\dot{t}^2+B(r)\dot{r}^2+ C(r) \dot{\phi}^2,
\end{equation}
where an overdot represents a derivative with respect to the affine
parameter. Since the Lagrangian is independent of $t$ and $\phi$, we
have two Killing vectors that result in two constants of motion,
\begin{equation}
p_t=\frac{\partial\mathcal{L}}{\partial\dot{t}}=-A(r)\dot{t}=-E~,~
p_\phi=\frac{\partial\mathcal{L}}{\partial\dot{\phi}}=C(r)\dot{\phi}=L,
\end{equation}
where $E$ and $L$ are, respectively, the energy and angular momentum
of the photon.  Using the null geodesics condition
$g_{\mu\nu}\dot{x}^\mu\dot{x}^\nu=0$, we obtain
\begin{equation}
AB\dot{r}^2+V_{eff}=E^2, \hspace{0.3cm}
V_{eff}={L^2}\frac{A(r)}{C(r)},
\label{Veffec}
\end{equation}
where $V_{eff}$ is the effective potential. A photon coming from a source at infinity may undergo a turning at some radius $r_0$ and escape to a faraway observer. At the turning point $r_0$, $\dot{r}=0$, i.e., $V_{eff}(r_0)=E^2$. This gives the following relationship
between the impact parameter $b$ ($=L/E$) (which remains constant throughout its trajectory) of the photon and the turning point $r_0$,
\begin{equation}
b^2=\frac{C(r_0)}{A(r_0)}.
\label{eq:impact_parameter}
\end{equation}
For such a photon which comes from a distant source, takes a turn at $r_0$ and escapes to a faraway observer, the deflection angle $\alpha(r_{0})$ can be obtained as
\begin{equation}
\alpha(r_{0})=I(r_{0})-\pi,
\label{eq:deflection1}
\end{equation}
where
\begin{equation}
I(r_{0})= 2\int^{\infty}_{r_{0}}\frac{dr}{\sqrt{\frac{R(r)C(r)}{B(r)}}}~,~
R(r)= \left(\frac{A_{0}C}{AC_{0}}-1 \right).
\label{eq:I1}
\end{equation}
We define the photon and the antiphoton sphere, respectively, as the locations of unstable and stable circular orbits (also known as light rings) of photons. Circular photon orbits satisfy $V_{eff}=E^2$ and $dV_{eff}/dr=0$, resulting in Eq. (\ref{eq:impact_parameter}) and
\begin{equation}
\frac{C'(r)}{C(r)}-\frac{A'(r)}{A(r)}=0~,
\label{eq:psph}
\end{equation}
respectively. In addition to the above equation, at the location of the photon and antiphoton sphere, we must have, respectively, $d^2V_{eff}/dr^2<0$ (maximum of the potential) and $d^2V_{eff}/dr^2>0$ (minimum of the potential). We denote the position of the photon sphere by $r = r_m$, and the corresponding critical impact parameter as $b = b_m=\sqrt{C(r_m)/A(r_m)}$. Equation (\ref{eq:psph}) is satisfied at $r=r_m$.

\section{Lensing of light in the strong deflection limit}
\label{sec:3}

We now study gravitational lensing in the strong deflection limit \cite{SL3}. We first review the known results 
when the turning point of light is outside the photon sphere, i.e., $r_0>r_m$. Such a situation arises in lensing from black holes (see Fig. \ref{fig1}) as well as in that from UCOs. 

\subsection{Strong bending of light due to a photon sphere}
\label{subsec:3a}

The strong gravitational lensing of light due to the presence of a photon sphere has been studied in \cite{SL3,Tsu1}. 
In such a case, a photon always takes a turn from outside the photon sphere ($r_0>r_m$); i.e., it always remains outside the photon sphere (see Fig. \ref{fig1}), 
and the strong deflection limit occurs when its impact parameter approaches the critical value $b_m$ from $b > b_m$.
In this case, the deflection angle in the strong deflection limit $r_{0}\rightarrow r_{m}$ or $b\rightarrow b_m$ is given by \cite{Tsu1}
\begin{equation}
\alpha(b)=-\bar{a}\log \left( \frac{b}{b_m}-1 \right) +\bar{b} +\mathcal{O}((b-b_m)\log(b-b_m)),
\label{eq:strong_alpha_1}
\end{equation}
where $\bar{a}$ and $\bar{b}$ are given by
\begin{equation}
\bar{a}=\sqrt{\frac{2B_{m}A_{m}}{C^{''}_{m}A_{m}-C_{m}A^{''}_{m}}}~,~
\bar{b}=\bar{a}\log \left[r^{2}_{m}\left(\frac{C_{m}^{''}}{C_{m}}-\frac{A_{m}^{''}}{A_{m}}\right)\right] +I_{R}(r_{m})-\pi,
\label{eq:strong_bbar_1}
\end{equation}
respectively, where the subscript $m$ implies that the corresponding quantities are evaluated at $r=r_m$. We now proceed to analyze situations where photons encounter an antiphoton sphere. 

\subsection{Strong bending of light experiencing an antiphoton sphere}
\label{subsec:3b}

This case arises in lensing from UCOs (not from black holes). Due to the presence of an antiphoton sphere, the height of the effective potential of a photon decreases from the photon 
sphere to a minimum value at the antiphoton sphere and starts rising again below this radius. 
In such cases, a photon with an impact parameter less than the critical value $b_m$ enters the photon sphere, 
passes through the antiphoton sphere, takes a turn at a radius inside the antiphoton sphere, and 
comes out of the photon sphere and escapes to a faraway observer (see Fig. \ref{fig2}). 
For such a photon, when the impact parameter approaches the critical value $b_m$ from $b < b_m$, 
it undergoes strong deflection. However, the strong deflection occurs when the photon on its trajectory is in the 
vicinity of the photon sphere. Therefore, to obtain the strong deflection formula in this case, we introduce a variable 
$z$ defined as 
\begin{equation}
z= 1-\frac{r_{m}}{r}.
\end{equation}
Putting this in $I(r_{0})$, we obtain
\begin{equation}
I(r_{0})=\int^{1}_{1-\frac{r_{m}}{r_0}}f(z,r_0,r_m)dz,
\end{equation}
where
\begin{equation}
f(z,r_0,r_m)= \frac{2r_{m}}{\sqrt{G(z,r_0,r_m)}}~,~
G(z,r_0,r_m)= R\frac{C}{B}(1-z)^{4}.
\label{defG}
\end{equation}
Since the strong deflection occurs around the photon sphere, we need to expand $G(z,r_0,r_m)$ 
around $r=r_m$ or $z=0$ to extract the divergent part. To this end, we first note that the expansions of a function 
$F(r)$ and its inverse $1/F(r)$ in the power of $z$ can be written as
\begin{eqnarray}
&~&F=F_m+F_m^{'}r_{m}z+\left(\frac{1}{2}F^{''}_{m}r_m^{2}+F_m^{'}r_m \right)z^{2}+\mathcal{O}(z^{3})\nonumber\\
&~&\frac{1}{F}
=\frac{1}{F_m}-\frac{F_m^{'}r_m}{F_{m}^{2}}z +\frac{r_{m}}{F_{m}^{3}} 
\left( -\frac{1}{2}F_{m}F_{m}^{''}r_{m} +F_{m}^{'2}r_{m}-F_{m}F_{m}^{'} \right)z^{2} +\mathcal{O}(z^{3}).\nonumber\\
\end{eqnarray}
Therefore, $R(r)$ can be expanded in the power of $z$ as
\begin{equation}
R(r)= \left(\frac{A_0 C_m}{C_0 A_m}-1\right) 
+\frac{r_{m}^2}{2} \frac{A_0 C_m}{C_0 A_m} \left(\frac{C_{m}^{''}}{C_{m}}-\frac{A_{m}^{''}}{A_{m}}\right)z^2 +\mathcal{O}(z^3). 
\end{equation}
Using similar expansion for the functions $B$ and $C$ in Eq. (\ref{defG}), 
we obtain the expansion of $G(z,r_0,r_m)$ in powers of $z$ as 
\begin{equation}
G(z,r_0,r_m)=\gamma+\delta z+\eta z^2+\mathcal{O}(z^3),
\end{equation}
where we have defined 
\begin{equation}
\gamma=\frac{C_m}{B_m}\left(\frac{A_0 C_m}{C_0 A_m}-1\right)
\label{gammaint}
\end{equation}
\begin{equation}
\delta=\frac{C_m}{B_m}\left(\frac{A_0 C_m}{C_0 A_m}-1\right)\left[-4+r_m\left(\frac{C_{m}^{'}}{C_{m}}-
\frac{B_{m}^{'}}{B_{m}} \right)\right]
\label{deltaint}
\end{equation}
\begin{eqnarray}
\eta &=&\frac{C_m}{B_m}\left(\frac{A_0 C_m}{C_0 A_m}-1\right)\left[6-r_m\left(3+\frac{B_m^{'}r_m}{B_m}\right)\left(\frac{C_{m}^{'}}{C_{m}}-\frac{B_{m}^{'}}{B_{m}} \right)\right. \nonumber \\
& & \left.  +\frac{r_m^2}{2}\left(\frac{C_{m}^{''}}{C_{m}}-\frac{B_{m}^{''}}{B_{m}}\right)\right]+
\frac{r_m^2}{2}\frac{C_m}{B_m}\frac{A_0 C_m}{C_0 A_m}\left(\frac{C_{m}^{''}}{C_{m}}-\frac{A_{m}^{''}}{A_{m}}\right)~.
\label{etaint}
\end{eqnarray}

Note that the heights of the effective potential (in units of the angular momentum squared) at the photon sphere $r=r_m$ and at $r=r_c$ are the same (see Fig. \ref{fig2}), 
i.e., $\frac{A(r_c)}{C(r_c)}=\frac{A(r_m)}{C(r_m)}$ or $b(r_c)=b_m$. Therefore, when the impact parameter approaches the critical value $b_m$ from $b<b_m$, the turning point $r_0$ approaches the radius $r_c$. Hence, in the limit 
$r_0 \rightarrow r_c$, $\left(\frac{A_0 C_m}{C_0 A_m}-1\right)\to 0$. In this limit, we also obtain
\begin{equation}
\gamma_m=\gamma\vert_{r_0=r_c}=0=\delta_m=\delta\vert_{r_0=r_c}
\end{equation}
and
\begin{equation}
\eta_m=\eta\vert_{r_0=r_c}=\frac{r_m^2}{2}\frac{C_m}{B_m}\left(\frac{C_{m}^{''}}{C_{m}}-\frac{A_{m}^{''}}{A_{m}}\right).
\end{equation}
Hence, we obtain
\begin{equation}
G_{m}(z)=\eta_m z^{2}+\mathcal{O}(z^{3}).
\end{equation}
This shows that the leading order of the divergence of $f(z, r_0,r_m)$ is $z^{-1}$ and that the integral $I(r_{0})$ diverges logarithmically in the strong deflection limit $r_0\rightarrow r_c$, as was the case for black holes in \cite{SL3}.

To extract the logarithmic divergence in the strong deflection limit, we split the integral $I(r_0)$ into a divergent part 
$I_D(r_0)$ and a regular part $I_R(r_0)$ so that $I(r_{0})=I_{D}(r_{0})+I_{R}(r_{0})$. 
The divergent part $I_D(r_0)$ is defined as
\begin{equation}
I_D(r_0)= \int^{1}_{1-\frac{r_m}{r_0}}f_D(z,r_0,r_m)dz~,~
f_D(z,r_0,r_m)
=\frac{2r_m}{\sqrt{\gamma+\delta z+\eta z^2}}.
\end{equation}
The regular part $I_R(r_0)$ is defined as
\begin{equation}
I_{R}(r_{0})=\int^{1}_{1-\frac{r_m}{r_0}} f_R(z,r_0,r_m)dz~,~
f_R(z,r_0,r_m)=f(z,r_0,r_m)-f_D(z,r_0,r_m).
\end{equation}
Integrating $I_D(r_0)$, we obtain 
\begin{equation}
I_D(r_0)=\frac{2r_m}{\sqrt{\eta}}\log \frac{\delta+2\eta+2\sqrt{\eta}\sqrt{\gamma+\delta+\eta}}{\delta+2\eta\left(1-\frac{r_m}{r_0}\right)+2\sqrt{\eta}\sqrt{\gamma+\delta\left(1-\frac{r_m}{r_0}\right)+\eta\left(1-\frac{r_m}{r_0}\right)^2}}.
\end{equation}
In the limit $r_0\to r_c$, treating $\gamma$ and $\delta$ as small parameters, we obtain after some algebra, 
\begin{equation}
I_D(r_0)=\frac{2r_m}{\sqrt{\eta_m}}\log\left[\frac{4\eta_m\left(\frac{r_m}{r_0}-1\right)}{\frac{C_m}{B_m}\left(\frac{A_0 C_m}{C_0 A_m}-1\right)}\right]+\mathcal{O}\left[\left(\frac{A_0 C_m}{C_0 A_m}-1\right)\log\left(\frac{A_0 C_m}{C_0 A_m}-1\right)\right].
\label{IDr0}
\end{equation}
Note that we can also write the expansion
\begin{eqnarray}
\frac{A_0 C_m}{C_0 A_m}&=&\frac{C_m}{A_m}\left[\frac{A_c+A_c'(r_0-r_c)+\mathcal{O}(r_0-r_c)^2}{C_c+C_c'(r_0-r_c)+\mathcal{O}(r_0-r_c)^2}\right]\nonumber\\
&=& 1+\left(\frac{A_c'}{A_c}-\frac{C_c'}{C_c}\right)(r_0-r_c)+\mathcal{O}(r_0-r_c)^2,
\label{rat}
\end{eqnarray}
where the subscript $c$ indicates that the quantities are evaluated at $r=r_c$, and we have used $\frac{A_c C_m}{C_c A_m}=1$ 
in the last equation. Using Eqs. (\ref{IDr0}) and (\ref{rat}), we obtain
\begin{eqnarray}
I_D(r_0)&=&-{\frac{2r_m}{\sqrt{\eta_m}}} \log \left( r_c-r_0 \right) +{\frac{2r_m}{\sqrt{\eta_m}}} \log \left[4\frac{B_m}{C_m}\left(\frac{r_m}{r_c}-1\right)\eta_m \left(\frac{C_c'}{C_c}-\frac{A_c'}{A_c}\right)^{-1}\right] \nonumber\\
&&+\mathcal{O}[(r_c-r_0)\log (r_c-r_0)].
\label{IDr0A}
\end{eqnarray}

It is more meaningful to write $I_D$ in terms of the impact parameter $b$. To this end, we first note that 
$b=b(r_0)=\frac{C(r_0)}{A(r_0)}$, $b_m=b(r_m)=\frac{C(r_m)}{A(r_m)}$ and hence, from Eq. (\ref{rat}), we obtain
\begin{equation}
r_0=r_c-\left(\frac{C_c'}{C_c}-\frac{A_c'}{A_c}\right)^{-1}\left( \frac{b_m^2}{b^2}-1 \right)
\label{r0}
\end{equation}
Therefore, from Eq. (\ref{IDr0A}), we obtain the divergent part $I_{D}(b)$ in the strong deflection limit $b\rightarrow b_m$ as 
\begin{eqnarray}
I_D(b)
&=&-{\frac{2r_m}{\sqrt{\eta_m}}} \log \left( \frac{b_m^2}{b^2}-1 \right) +{\frac{2r_m}{\sqrt{\eta_m}}} \log \left[4\frac{B_m}{C_m}\left(\frac{r_m}{r_c}-1\right)\eta_m\right] \nonumber\\
&&+\mathcal{O}[(b_m^2-b^2)\log (b_m^2-b^2)].
\end{eqnarray}

In the strong deflection limit $r_0\rightarrow r_c$ or $b\rightarrow b_{m}$ (keep in mind that $b_c=b_m$), we now expand the regular 
part $I_R(r_0)$ in powers of $r_c-r_0$ and keep the leading order term which can be integrated analytically or 
numerically. We find that 
\begin{equation}
I_R(r_0)= \int^{1}_{1-\frac{r_m}{r_c}} f_R(z,r_c,r_m)dz+\mathcal{O}((r_c-r_0)\log(r_c-r_0))
\end{equation}
which can be expressed in terms of the impact parameter as 
\begin{equation}
I_R(b)= \int^{1}_{1-\frac{r_m}{r_c}} f_R(z,b_m)dz+\mathcal{O}((b_m^2-b^2)\log(b_m^2-b^2)).
\end{equation}
Finally, the deflection angle in the strong deflection limit $r_0\rightarrow r_c$ or $b\rightarrow b_m$ can be written as
\begin{equation}
\alpha(b)=-\bar{a}\log \left( \frac{b_m^2}{b^2}-1 \right) +\bar{b} +\mathcal{O}((b_m^2-b^2)\log(b_m^2-b^2)),
\label{eq:strong_alpha_2}
\end{equation}
where
\begin{equation}
\bar{a}=2\sqrt{\frac{2B_m A_m}{C^{''}_m A_m-C_m A^{''}_m}},
\label{eq:strong_abar_2}
\end{equation}
\begin{equation}
\bar{b}=\bar{a}\log \left[2r_m^2\left(\frac{C_m^{''}}{C_m}-\frac{A_m^{''}}{A_m}\right)\left(\frac{r_m}{r_c}-1\right)\right] +I_R(r_c)-\pi.
\label{eq:strong_bbar_2}
\end{equation}
Note that Eqs. (\ref{eq:strong_alpha_2})--(\ref{eq:strong_bbar_2}) obtained for the relativistic images formed inside the 
photon sphere are completely different from Eqs. (\ref{eq:strong_alpha_1}) and (\ref{eq:strong_bbar_1}) obtained in \cite{Tsu1} 
for those formed outside the photon sphere. Especially we see that $\bar{a}$ in this case is twice that of the earlier case,
and the expression for $\bar{b}$ contains the factor $\left(\frac{r_m}{r_c}-1\right)$ which is absent in the earlier case. 
This implies that the bending angle for the inner relativistic images starts diverging much before (in terms of the 
difference $|b_m-b|$ in the impact parameter) that for the outer relativistic images as the critical impact parameter is reached.
 As a result, the angular separation between the inner images is much more than that of the outer images.
 
Here, we point out that one can use the approximation $\left( \frac{b_m^2}{b^2}-1 \right)=\left( \frac{b_m}{b}+1 \right)\left( \frac{b_m}{b}-1 \right) \simeq 2\left( \frac{b_m}{b}-1 \right)$ in Eq. (\ref{eq:strong_alpha_2}) as is done for the images outside the photon sphere [see Eq. (\ref{eq:strong_alpha_1})]. However, doing so will introduce greater error in the results as the difference between the critical impact parameter $b_m$ and the impact parameters $b$ at which the inner images are formed, i.e., $(b_m-b)$, is relatively larger than that of the outer images. Therefore, we do not use this approximation for the inner images.

\section{Observables in gravitational lensing}
\label{sec:4}

Having elaborated upon the necessary formalism, we are now in a position to obtain analytic expressions of various observables (commonly used in the literature of gravitational lensing) for the relativistic images formed inside the photon sphere. Analytic expressions of these observables for the relativistic images formed outside the photon sphere are obtained in \cite{SL3}. We closely follow \cite{SL3} to obtain the corresponding expressions for the inner images.

We start from the lens equation in the strong field limit (thin lens approximation), which is given by
\begin{equation}
\beta=\theta-\frac{D_{LS}}{D_{OS}}\Delta\alpha_n,
\end{equation}
where $D_{LS}$ is the distance between the lens and the source, $D_{OS}$ is the distance between the observer and the 
source, $D_{OS}=D_{OL}+D_{LS}$, $D_{OL}$ is the distance between the observer and the lens, $\beta$ is the angular 
separation between the source and the lens, $\theta$ is the angular separation between the lens and the image, and $\Delta\alpha_n=\alpha(\theta)-2\pi n$ is the offset of the deflection angle after subtracting all the winding (encoded in $n$) 
undergone by the photon.

The angular position $\theta_n^0$ and the magnification $\mu_n$ of the $n$th relativistic image formed outside the photon sphere 
are, respectively, given  by \cite{SL3}
\begin{equation}
\theta_n^0=\frac{b_m}{D_{OL}}(1+e_n)=\theta_{\infty}(1+e_n), \quad e_n=e^{\frac{\bar{b}-2n\pi}{\bar{a}}},
\end{equation}
\begin{equation}
\mu_n = \frac{b_m^2 D_{OS}e_n(1+e_n)}{\bar{a}\beta D_{OL}^2 D_{LS}},
\end{equation}
where $\theta_{\infty}=b_m/D_{OL}$ is the angular position of the relativistic image formed at the photon sphere. Note that the angular positions of the images decreases with $n$, implying that, in the outer image system, the first relativistic image is the outermost one and the image with the angular position $\theta_{\infty}$ is the innermost one. It is assumed that only the outermost (first) image of the outer images can be resolved from the rest. 
Therefore, we can define two more observables, namely the angular separation $s_1$ between the first image and 
the rest and the ratio $r_1$ between the flux of the first image and the total flux of all the other images. These are given by \cite{SL3}
\begin{equation}
s_1=\theta_1-\theta_\infty,
\end{equation}
\begin{equation}
r_1=\frac{\mu_1}{\sum_{m=2}^\infty \mu_m}.
\label{eq:outer_flux_ratio}
\end{equation}

Let us now obtain the above observables for the relativistic images formed inside the photon sphere, i.e., for those formed 
due to the presence of an antiphoton sphere. To this end, we first  note that if $b$ is the impact parameter at which the $n$th relativistic image is formed, then we can write $\theta=b/D_{OL}$. Therefore, in terms of $\theta$, the deflection angle (\ref{eq:strong_alpha_2}) can be written as
\begin{equation}
\alpha(\theta)=-\bar{a}\log \left( \frac{b_m^2}{D_{OL}^2\theta^2}-1 \right) +\bar{b} +\mathcal{O}((b_m^2-D_{OL}^2\theta^2)\log(b_m^2-D_{OL}^2\theta^2)).
\end{equation}
The observables for inner images are denoted by the subscript $-n$. This $-$ sign before $n$ indicates that we are talking about the $n$th relativistic image of the inner images. 
Also, we replace $\Delta\alpha_n$ in that we have introduced before by $\Delta\alpha_{-n}$. To obtain the 
offset $\Delta\alpha_{-n}$, we expand $\alpha(\theta)$ around $\theta=\theta_{-n}^0$, where $\alpha(\theta_{-n}^0)=2\pi n$. 
Using $\alpha(\theta_{-n}^0)=2\pi n$, we obtain
\begin{equation}
\theta_{-n}^0=\frac{b_m}{D_{OL}}\frac{1}{\sqrt{1+e_{-n}}}=\frac{\theta_{-\infty}}{\sqrt{1+e_{-n}}}, \quad e_{-n}=e^{\frac{\bar{b}-2n\pi}{\bar{a}}},
\end{equation}
where $\theta_{-\infty}=b_m/D_{OL}$ is the angular position of the relativistic image formed at the photon sphere. Note that in contrast to that for the images formed outside the photon sphere, the angular positions of the images formed inside the photon sphere increase with $n$, implying that, in the inner image system, the first relativistic image is the innermost one and the image with the angular position $\theta_{-\infty}$ is the outermost one. Note also that $\theta_{\infty}=\theta_{-\infty}$. Now, defining $\Delta\theta_{-n}=\theta-\theta_{-n}^0$, we obtain
\begin{equation}
\alpha(\theta) \simeq \alpha(\theta_{-n}^0)+\frac{d\alpha}{d\theta}\Big\vert_{\theta_{-n}^0}\Delta\theta_{-n}.
\label{dalpha}
\end{equation}
Using Eq.(\ref{dalpha}), we obtain
\begin{equation}
\Delta\alpha_{-n}=\frac{2\bar{a}D_{OL}}{e_{-n} b_m}(1+e_{-n})^{3/2}\Delta\theta_{-n}.
\end{equation}
With this, the lens equation becomes
\begin{equation}
\beta=\theta_{-n}^0+\Delta\theta_{-n}-\frac{D_{LS}}{D_{OS}}\frac{2\bar{a}D_{OL}}{e_{-n} b_m}(1+e_{-n})^{3/2}\Delta\theta_{-n}.
\label{lenseq}
\end{equation}
The second term in the above equation is negligible compared to the last one since $b_m\ll D_{OL}$. Neglecting this 
second term, the angular position of the relativistic images is given by
\begin{equation}
\theta=\theta_{-n}^0-\frac{b_m e_{-n} D_{OS}}{2\bar{a}D_{LS}D_{OL}}\frac{(\beta-\theta_{-n}^0)}{(1+e_{-n})^{3/2}}.
\end{equation}
Note that the correction to $\theta_{-n}^0$ is negligible compared to $\theta_{-n}^0$. Therefore, we approximate 
the position of the images by $\theta_{-n}^0$ in order to calculate the magnifications of the images given by
\begin{equation}
\mu_{-n}=\frac{1}{(\beta/\theta)(\partial \beta/\partial \theta)}\Big\vert_{\theta_{-n}^0}.
\end{equation}
Now we have from Eq.(\ref{lenseq}),
\begin{equation}
\frac{\partial \beta}{\partial \theta}\Big\vert_{\theta_{-n}^0} = 1-\frac{D_{LS}}{D_{OS}}\frac{2\bar{a}D_{OL}}{e_{-n} b_m}(1+e_{-n})^{3/2} 
\simeq -\frac{D_{LS}}{D_{OS}}\frac{2\bar{a}D_{OL}}{e_{-n} b_m}(1+e_{-n})^{3/2},
\end{equation}
where we have neglected the first term as it is negligible compared to the second term (because $b_m\ll D_{OL}$). 
Therefore, the magnification becomes
\begin{equation}
\mu_{-n} = -\frac{b_m^2 D_{OS}}{2\bar{a}\beta D_{OL}^2 D_{LS}}\frac{e_{-n}}{(1+e_{-n})^2}.
\end{equation}
Beside the angular positions and magnifications of the relativistic images, we define two other observables, 
namely the angular separation $s_{-n}$ between the $n$th and $(n+1)$th images and the ratio $r_{-n}$ between the flux of each of the first three images and the total flux of all the other images. Thus
\begin{equation}
s_{-n}=\vert\theta_{-n}-\theta_{-(n+1)}\vert,
\end{equation}
\begin{equation}
r_{-n}=\frac{|\mu_{-n}|}{\sum_{m=2}^\infty \mu_m+\sum_{m=4}^\infty |\mu_{-m}|} \quad (n=1,2,3).
\end{equation}
Note that, in the absence of the images inside the photon sphere (i.e. in the case of black holes), the flux ratio for the first image of the images outside the photon sphere is given by Eq. (\ref{eq:outer_flux_ratio}). However, in the presence of the inner images (i.e. in the case of UCOs), we define the same by
\begin{equation}
r_{1}=\frac{\mu_1}{\sum_{m=2}^\infty \mu_m+\sum_{m=4}^\infty |\mu_{-m}|}.
\end{equation}
Our results for these variables for different geometries is presented in Table \ref{Table1}. Here we have restored $G$ and $c$ by replacing $M$ by $(GM)/c^2$. Here, the mass $M$ and the distance $D_{OL}$ are taken to be those of the supermassive black hole Sgr A$^*$ at the center of our Galaxy.
\begin{table}[h!]
\centering
\caption{The angles are in microarc sec and $r_{n/-n}$ is converted to magnitude using $\mathcal{R}_{n/-n}=2.5\log r_{n/-n}$. Here, we have taken $M=4.31\times 10^6 M_{\odot}$, $D_{OL}=D_{LS}=7.86$ Kpc, which are the parameters for the supermassive black hole Sgr A$^*$ at center of our Galaxy and $\beta=5^\circ$.}
 \begin{tabular}{| c | c | c | c | c | c | c |} 
 \hline\hline
  & Schwarzschild & \multicolumn{2}{c|}{Schwarzschild interior} & Interior & RN naked & Noncommutative \\
  & black hole & \multicolumn{2}{c|}{(Synge)} & (Florides) & singularity & Schwarzschild \\
  & & $R=2.5 M$ & $R=2.7 M$ & $R=2.5 M$ & $\frac{Q^2}{M^2}=1.05$ & $\frac{\sqrt{\theta}}{M}=0.6$ \\
 \hline
  $\theta_1$ & 28.2802 & 28.2802 & 28.2802 & 28.2802 & 21.3541 & 28.1010 \\
  $\theta_{\infty}$ & 28.2449 & 28.2449 & 28.2449 & 28.2449 & 21.1592 & 28.0282 \\
  $\theta_{-3}$ & --- & 28.2353 & 28.2395 & 28.2404 & 21.0032 & 27.9864 \\
  $\theta_{-2}$ & --- & 28.0251 & 28.1218 & 28.1419 & 20.0692 & 27.5656 \\
  $\theta_{-1}$ & --- & 24.1813 & 25.7527 & 26.1160 & 15.6203 & 23.8330 \\
  $\mu_1\times 10^{22}$ & 5.3850 & 5.3850 & 5.3850 & 5.3850 & 14.3633 & 8.5121 \\
  $\mu_{-3}\times 10^{22}$ & --- & $-0.7300$ & $-0.4068$ & $-0.3402$ & $-5.5914$ & $-2.4290$ \\
  $\mu_{-2}\times 10^{22}$ & --- & $-16.3947$ & $-9.2574$ & $-7.7632$ & $-34.8801$ & $-25.8732$ \\
  $\mu_{-1}\times 10^{22}$ & --- & $-210.285$ & $-150.655$ & $-133.240$ & $-95.7851$ & $-163.620$ \\
  $s_1$ & 0.0353 & 0.0353 & 0.0353 & 0.0353 & 0.1949 & 0.0728 \\
  $s_{-3}$ & --- & 0.0092 & 0.0051 & 0.0043 & 0.1350 & 0.0381 \\
  $s_{-2}$ & --- & 0.2102 & 0.1177 & 0.0985 & 0.9340 & 0.4208 \\
  $s_{-1}$ & --- & 3.8438 & 2.3691 & 2.0259 & 4.4489 & 3.7326 \\
  $\mathcal{R}_1$ & 15.71 & 12.07 & 13.11 & 13.39 & 6.33 & 8.33 \\
  $\mathcal{R}_{-3}$ & --- & 7.08 & 6.65 & 6.48 & 3.97 & 5.20 \\
  $\mathcal{R}_{-2}$ & --- & 14.85 & 14.46 & 14.30 & 8.54 & 11.11 \\
  $\mathcal{R}_{-1}$ & --- & 21.23 & 21.44 & 21.41 & 11.07 & 15.72 \\
 \hline\hline
 \end{tabular}
\label{Table1}
\end{table}

The contents of Table \ref{Table1} are now summarized. 
\begin{itemize}
\item
The second column
is the results for the Schwarzschild black hole with the ADM mass $M$. 
\item
The third and fourth columns are the results for the interior Schwarzschild solution due to Synge \cite{Synge}, with 
the corresponding quantities in Eq. (\ref{genmetric}) being 
\begin{equation}
A = \left(\frac{3}{2} \sqrt{1-\frac{2 M}{R}}-\frac{1}{2} \sqrt{1-\frac{2 Mr^2}{R^3}}\right)^2~,~
B = \left(1-\frac{2Mr^2}{R^3}\right)^{-1}~,~C = r^2~.
\label{eq:synge}
\end{equation}
Here, $r=R$ denotes a hypersurface across which the metric is matched to an external Schwarzschild solution
with ADM mass $M$, where we have taken $R/M=2.5$ and $2.7$.
\item
The fifth column is the results for an interior
Schwarzschild solution due to Florides \cite{Florides} where the metric components of Eq. (\ref{genmetric}) are given as
\begin{equation}
A = \frac{\left(1 - 2M/R\right)^{\frac{3}{2}}}{\left(1-2Mr^2/R^3\right)^{1/2}}~,~B = \left(1 - \frac{2Mr^2}{R^3}\right)^{-1}~,~C = r^2~,
\end{equation}
Here, $r=R$ denotes a hypersurface across which the metric is matched to an external Schwarzschild solution
with ADM mass $M$, where we have taken $R/M=2.5$. Note that for this solution, the radial pressure vanishes and thus the Florides solution might be thought of as the
geometry of an Einstein cluster. Here, it should be thought of as a toy model for an anisotropic UCO.
\item
The sixth column in Table \ref{Table1} shows the results for the Reissner-Nordstrom naked singularty where
we have taken the (square of the) charge to mass ratio to be $1.05$.
\item
The seventh column is the results for a noncommutative
Schwarzschild regular solution \cite{noncommutative} where the metric components of Eq. (\ref{genmetric}) are given as
\begin{equation}
A = B^{-1}= 1-\frac{4M}{\sqrt{\pi}r}\gamma(3/2,r^2/4\theta),~C = r^2~,
\end{equation}
where $\gamma(3/2,r^2/4\theta)$ is the lower incomplete gamma function,
\begin{equation}
\gamma(3/2,r^2/4\theta)= \int_0^{r^2/4\theta}t^{1/2}e^{-t}dt.
\end{equation}
Here, we have considered the horizonless case with $\sqrt{\theta}/M=0.6$.
 
\end{itemize}

\section{Discussions and conclusions}
\label{sec:5}

We study strong gravitational lensing and formation of relativistic images by horizonless ultracompact objects and compare our results with that of black holes. It is well known that, for black holes, relativistic images are formed only outside their photon spheres. In contrast, for horizonless ultracompact objects, additional relativistic images can form inside their photon sphere radius. In this paper, we provide an analytical approach to deal with strong gravitation lensing from such ultracompact objects, which is substantially different from the black hole cases, first reported
by Bozza. We obtain an analytic expression for the bending angle as well as lensing observables such as angular positions, angular separations and magnifications of relativistic images formed inside the photon sphere and compare them with those of images formed outside it. We find that both the angular separation and magnification of the images initially decrease from the outermost (first) image outside the photon sphere, become minimum at the photon sphere and again start increasing becoming maximum for the innermost (first) image inside the photon sphere. Significantly, we see also that the magnification and separation of the relativistic images that appear inside the photon sphere radius are much larger than the corresponding images outside this radius. In fact, the angular separation between the first two
images inside the photon sphere radius can be $2$ orders of magnitude more than that of the first two images outside this
radius for the interior Schwarzschild solution, and the magnification of the first image inside this radius is about $40$ times
the one outside it. This result indicates that it might be easier to detect possible images inside the photon sphere compared
to the ones outside it.

\begin{figure}[h]
\centering
\centerline{\subfigure[]{\includegraphics[scale=0.8]{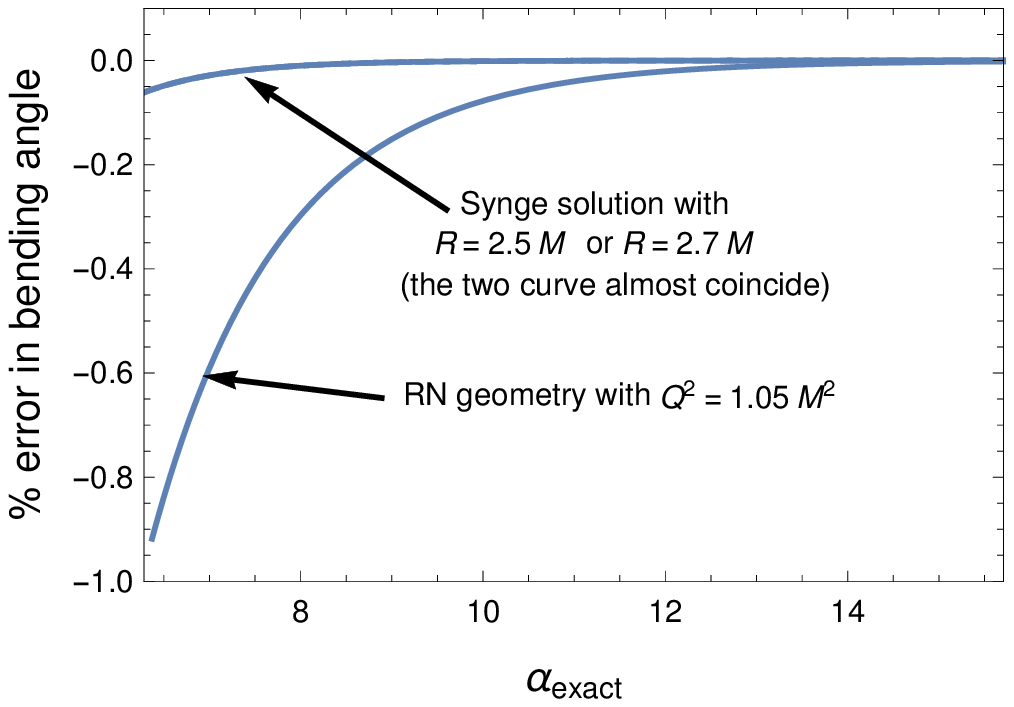}}
\subfigure[]{\includegraphics[scale=0.75]{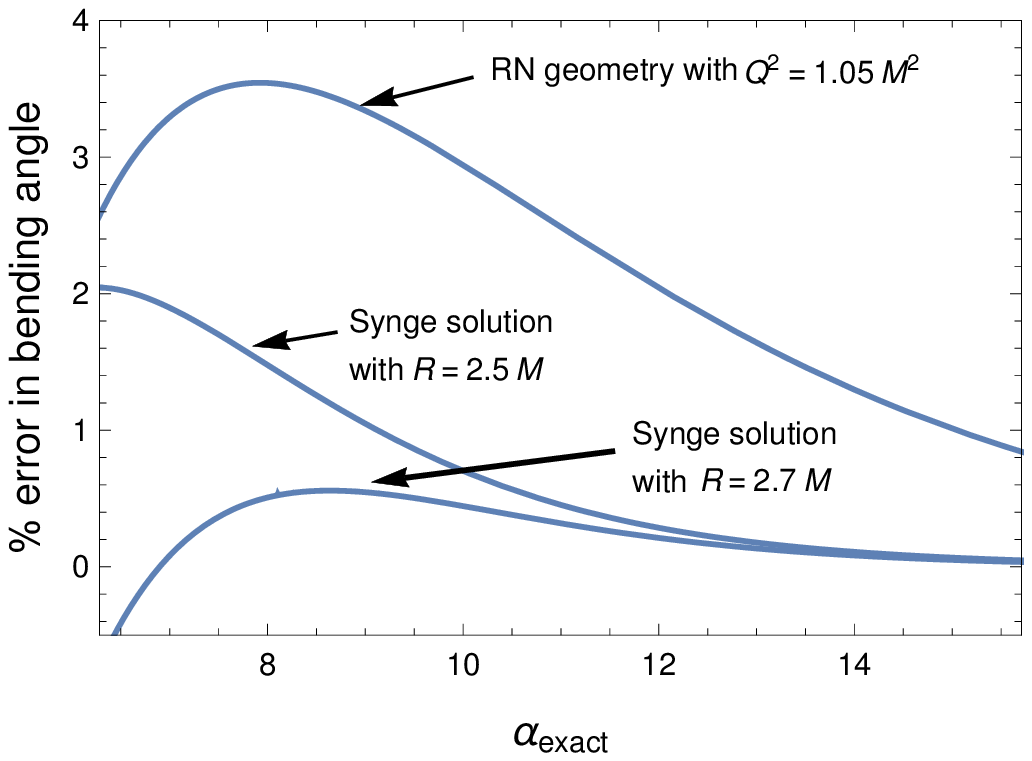}}}
\caption{$\%$ error $(\frac{\alpha-\alpha_{exact}}{\alpha_{exact}}\times 100)$ in bending angle as a function of $\alpha_{exact}$ in the strong deflection limit $(\alpha\geq 2\pi)$ for images $(a)$ outside and $(b)$ inside the photon sphere. Here, $\alpha_{exact}$ is obtained by numerical integration of Eq. (\ref{eq:deflection1}) and $\alpha$ is the analytic expression obtained in Eqs. (\ref{eq:strong_alpha_1}) and (\ref{eq:strong_alpha_2}) for the images outside and inside the photon sphere, respectively.} 
\label{error}
\end{figure}

Overall, the picture that emerges is as follows. For lensing by black holes, one would obtain closely separated images outside
the radius of the photon sphere. For UCOs on the other hand, one expects to see relatively wide separation between 
images up to the photon sphere, and more closely separated ones beyond this radius. This is a distinctive feature that,
if detected, can distinguish between images from black holes and UCOs. As an ending note, we show the percentage error obtained in the bending angle in the strong deflection limit. Note that the percentage error for the images inside the photon sphere is relatively larger than that for the images outside it, as shown in Fig. \ref{error}. This is because the difference between the critical impact parameter $b_m$ and the impact parameters $b$ at which the inner images are formed, i.e., $(b_m-b)$ is relatively larger than that of the outer images.

\begin{thebibliography}{99}
\bibitem{Weinberg}
S. Weinberg, {\em Gravitation And Cosmology} (John Wiley \& Sons, New York, 1972). 

\bibitem{BozzaReview}
V.~Bozza,
  Gravitational Lensing by Black Holes,
  Gen.\ Relativ.\ Gravit.\  {\bf 42}, 2269 (2010).

\bibitem{EHT}
The Event Horizon Telescope, https://eventhorizontelescope.org/\\
See also D. Castelvecchi, How to hunt for a black hole with a telescope the size of Earth, 
Nat. News {\bf 543}, 478 (2017).

\bibitem{SL1} 
K. S. Virbhadra and G. F. R. Ellis, Schwarzschild black hole lensing, Phys. Rev. D {\bf 62}, 084003 (2000).

\bibitem{SL2} 
V. Bozza, S. Capozziello, G. Iovane, and G. Scarptta, Strong field limit of black hole gravitational lensing, Gen. Relativ. Gravit. {\bf 33}, 1535 (2001).

\bibitem{SL3} 
V. Bozza, Gravitational lensing in the strong field limit, Phys. Rev. D {\bf 66}, 103001 (2002).


\bibitem{UCO1} V. Cardoso, E. Franzin, and P. Pani, Is the Gravitational-Wave Ringdown a Probe of the Event Horizon?, Phys. Rev. Lett. {\bf 116}, 171101 (2016); Erratum, Phys. Rev. Lett. {\bf 117}, 089902 (2016).

\bibitem{UCO2} P. V. P. Cunha, C. A. R. Herdeiro, and M. J. Rodriguez, Does the black hole shadow probe the event horizon geometry?, Phys. Rev. D {\bf 97}, 084020 (2018).

\bibitem{UCO3} R. A. Konoplya, Z. Stuchlik, and A. Zhidenko, Echoes of compact objects: New physics near the surface and matter at a distance, Phys. Rev. D {\bf 99}, 024007 (2019).

\bibitem{UCO4} A. Urbano and H. Veermae, On gravitational echoes from ultracompact exotic stars, J. Cosmol. Astropart. Phys. 04 (2019) 011.

\bibitem{UCO5} R. Carballo-Rubio, F. Di Filippo, S. Liberati, and M. Visser, Phenomenological aspects of black holes beyond general relativity, Phys. Rev. D {\bf 98}, 124009 (2018).

\bibitem{WL1} J. G. Cramer, R. L. Forward, M. S. Morris, M. Visser, G. Benford, and G. A. Landis, Natural wormholes as gravitational lenses, Phys.Rev. D {\bf 51}, 3117 (1995).

\bibitem{WL2} M. Safonova, D. F. Torres, and G. E. Romero, Microlensing by natural wormholes: Theory and simulations, Phys. Rev. D {\bf 65}, 023001 (2001).

\bibitem{WL3} K. K. Nandi, Y. -Z. Zhang, and A. V. Zakharov, Gravitational lensing by wormholes, Phys. Rev. D {\bf 74}, 024020 (2006).

\bibitem{WL4} F. Abe, Gravitational microlensing by the Ellis wormhole, Astrophys. J. {\bf 725}, 787 (2010).

\bibitem{WL5} K. Nakajima and H. Asada, Deflection angle of light in an Ellis wormhole geometry, Phys. Rev. D {\bf 85}, 107501 (2012).

\bibitem{WL6} N. Tsukamoto, T. Harada, and K. Yajima, Can we distinguish between black holes and wormholes by their Einstein-ring systems?, Phys. Rev. D {\bf 86}, 104062 (2012).

\bibitem{WL7} C. Bambi, Can the supermassive objects at the centers of galaxies be traversable wormholes? The first test of strong gravity for mm/sub-mm very long baseline interferometry facilities, Phys. Rev. D {\bf 87}, 107501 (2013).

\bibitem{WL8} P. G. Nedkova, V. Tinchev, and S. S. Yazadjiev, Shadow of a rotating traversable wormhole, Phys. Rev. D {\bf 88}, 124019 (2013).

\bibitem{WL9} T. Ohgami and N. Sakai, Wormhole shadows, Phys. Rev. D {\bf 91}, 124020 (2015).

\bibitem{WL10}  A. Abdujabbarov, B. Juraev, B. Ahmedov, and Z. Stuchlik, Shadow of rotating wormhole in plasma environment, Astrophys. Space Sci. {\bf 361}, 226 (2016).

\bibitem{WL11} N. Tsukamoto, Strong deflection limit analysis and gravitational lensing of an Ellis wormhole, Phys. Rev. D {\bf 94}, 124001 (2016).

\bibitem{WL12} N. Tsukamoto and T. Harada, Light curves of light rays passing through a wormhole, Phys. Rev. D {\bf 95}, 024030 (2017).

\bibitem{WL13} R. Shaikh and S. Kar, Gravitational lensing by scalar-tensor wormholes and the energy conditions, Phys. Rev. D {\bf 96}, 044037 (2017).

\bibitem{WL14} K. K. Nandi, R. N. Izmailov, A. A. Yanbekov, and A. A. Shayakhmetov, Ring-down gravitational waves and lensing observables: How far can a wormhole mimic those of a black hole?, Phys. Rev. D {\bf 95}, 104011 (2017).

\bibitem{WL15} K. Jusufi and A. Ovgun, Gravitational lensing by rotating wormholes, Phys. Rev. D {\bf 97}, 024042 (2018).

\bibitem{WL16} R. Shaikh, Shadows of rotating wormholes, Phys. Rev. D {\bf 98}, 024044 (2018).

\bibitem{WL17} A. Ovgun, Light deflection by Damour-Solodukhin wormholes and Gauss-Bonnet theorem, Phys. Rev. D {\bf 98}, 044033 (2018).

\bibitem{WL18} K. Jusufi, N. Sarkar, F. Rahaman, A. Banerjee, and S. Hansraj, Deflection of light by black holes and massless wormholes in massive gravity, Eur. Phys. J. C {\bf 78}, 349 (2018).

\bibitem{WL19} K. K. Nandi, R. N. Izmailov, E. R. Zhdanov, and A. Bhattacharya, Strong field lensing by Damour-Solodukhin wormhole, J. Cosmol. Astropart. Phys. 07 (2018) 027.

\bibitem{WL20} G. Gyulchev, P. Nedkova, V. Tinchev, and S. Yazadjiev, On the shadow of rotating traversable wormholes, Eur. Phys. J. C {\bf 78}, 544 (2018).

\bibitem{WL21} M. Amir, K. Jusufi, A. Banerjee, and S. Hansraj, Shadow images of Kerr-like wormholes, arXiv:1806.07782.

\bibitem{WL22} R. Shaikh, P. Banerjee, S. Paul, and T. Sarkar, A novel gravitational lensing feature by wormholes, Phys. Lett. B {\bf 789}, 270 (2019).

\bibitem{NL1} K. S. Virbhadra, D. Narasimha, and S. M. Chitre, Role of the scalar field in gravitational lensing, Astron. Astrophys. {\bf 337}, 1 (1998).

\bibitem{NL2} K. S. Virbhadra and G. F. R. Ellis, Gravitational lensing by naked singularities, Phys. Rev. D {\bf 65}, 103004 (2002).

\bibitem{NL3} K. S. Virbhadra and C. R. Keeton, Time delay and magnification centroid due to gravitational lensing by black holes and naked singularities, Phys. Rev. D {\bf 77}, 124014 (2008).

\bibitem{NL4} G. N. Gyulchev and S. S. Yazadjiev, Gravitational lensing by rotating naked singularities, Phys. Rev. D {\bf 78}, 083004 (2008).

\bibitem{NL5} S. Sahu, M. Patil, D. Narasimha, and P. S. Joshi, Can strong gravitational lensing distinguish naked singularities from black holes?,  Phys. Rev. D {\bf 86}, 063010 (2012).

\bibitem{NL6} D. Dey, K. Bhattacharya, and T. Sarkar, Astrophysics of Bertrand space-times," Phys. Rev. D {\bf 88}, 083532 (2013).

\bibitem{NL7} R. Shaikh, P. Kocherlakota, R. Narayan, and P. S. Joshi, Shadows of spherically symmetric black holes and naked singularities, Mon. Not. R. Astron. Soc. {\bf 482}, 52 (2019).

\bibitem{NL8} P. Banerjee, S. Paul, and T. Sarkar, On strong gravitational lensing in rotating galactic space-times," arXiv:1804.07030.

\bibitem{Cunha1} P. V. P. Cunha, J. A. Font, C. Herdeiro, E. Radu, N. Sanchis-Gual, and M. Zilhao, Lensing and dynamics of ultracompact bosonic stars, Phys. Rev. D {\bf 96}, 104040 (2017).

\bibitem{CO} H. Chakrabarty, A. B. Abdikamalov, A. A. Abdujabbarov, and C. Bambi, Weak gravitational lensing: A compact object with arbitrary quadrupole moment immersed in plasma, Phys. Rev. D {\bf 98}, 024022 (2018).

\bibitem{gravastar} T. Kubo and N. Sakai, Gravitational lensing by gravastars, Phys. Rev. D {\bf 93}, 084051 (2016).

\bibitem{Kunha}
P.~V.~P.~Cunha, E.~Berti, and C.~A.~R.~Herdeiro,
  Light-Ring Stability for Ultracompact Objects,
  Phys.\ Rev.\ Lett.\  {\bf 119}, 251102 (2017).

\bibitem{Hod}
S.~Hod, `On the number of light rings in curved space-times of ultra-compact objects,
Phys.\ Lett.\ B {\bf 776}, 1 (2018).

\bibitem{Cunha2}
P.~V.~P.~Cunha and C.~A.~R.~Herdeiro,
  Shadows and strong gravitational lensing: a brief review,
  Gen.\ Relativ.\ Gravit.\  {\bf 50}, 42 (2018).

\bibitem{Tsu1}
  N.~Tsukamoto,
  Deflection angle in the strong deflection limit in a general asymptotically flat, static, spherically symmetric space-time,
  Phys.\ Rev.\ D {\bf 95}, 064035 (2017).
  
\bibitem{mandar} M. Patil, P. Mishra, and D. Narasimha, Curious case of gravitational lensing by binary black holes: A tale of two photon spheres, new relativistic images and caustics, Phys. Rev. D {\bf 95}, 024026 (2017).

\bibitem{Synge}
  J.~L.~Synge,
  {\em Relativity: The General Theory},
  (North-Holland, Amsterdam, 1960).

\bibitem{Florides}
P. S. Florides, A new interior Schwarzschild solution,
Proc. R. Soc. A {\bf 337}, 529 (1974).

\bibitem{noncommutative} P.~Nicolini, A.~Smailagic, and E.~Spallucci,
  Noncommutative geometry inspired Schwarzschild black hole,
  Phys.\ Lett.\ B {\bf 632}, 547 (2006).
  
\end{thebibliography}
\end{document}